\documentclass{article}

\usepackage{graphicx}      % include this line if your document contains figures
\usepackage[square,numbers,sort&compress]{natbib}
\usepackage{amsfonts}
\usepackage{amsmath}
\usepackage{amssymb}
\usepackage{graphicx}
\usepackage{mathptmx}
\usepackage{tabularx}
\usepackage{boxedminipage}
\usepackage{subcaption}
\usepackage{geometry}
\begin{document}

\begin{center}
\LARGE {Utilization of G-Programming Language for Educational Control Application: Case Study of Magnetic Levitation of Elastic Beam} 
\end{center}
\vspace{0.5cm} 
\begin{center} 
\Large {*Abdallah Amr *Mostafa Eshra *Ayman A. Nada} \\[3pt] 
\normalsize {School of Innovative Design Engineering, Faculty of Engineering, Egypt-Japan University of Science and Technology E-JUST, Alexandria 21934, Egypt. email:{abdallah.amr@ejust.edu.eg}} \\ [1cm] 

\end{center}
\begin{center}

\begin{abstract}
This paper presents the practical employment of G-Programming tools to demonstrate, design, and implement traditional control algorithms upon magnetic levitation system. The complexity of controlling this type of fast dynamic and sensitive system is vital for highlighting the capabilities of LabVIEW G-programming in control education.  PID and Lead-Lag controllers are designed and implemented within the LabVIEW environment, with the ability to tune and optimize the controllers utilizing the Virtual Instruments (VIs) of the control design and simulation toolkit. The paper enables the reader to understand the modelling, testing the control action, and dynamic simulation of the system. Then, deploying the control law in real time. It can be concluded that the G-programming shows a suitable and easy tool for facilitating hands-on, experiential learning and validation in control systems engineering.
\end{abstract}

\noindent\rule{12cm}{0.4pt}
\end{center}

%===============================================================================

\section{Introduction}

Control systems engineering lies at the heart of various modern technologies, encompassing everything from industrial automation to consumer robotics. Educating young engineers in this vital field necessitates practical tools that bridge the gap between theoretical concepts and real-time applications which is the main topic of this paper. Specifically, the paper presents the ability of LabVIEW's G-programming language as an efficient, interactive, and friendly-use tool for control education.
Several factors contribute to making LabVIEW a good tool for teaching control engineering, as presented in \cite{Wang2010, Cruz2016, nada2024development}; the real-time Application focus as LabVIEW excels in facilitating real-time control system design and experimentation as that in \cite{bourhnane2019real}. Its inherent capabilities in data acquisition, hardware interfacing, and code execution seamlessly integrate theoretical frameworks with understandable outcomes. Educational and Theoretical Alignment in LabVIEW's graphical programming environment mirrors the flowcharts and block diagrams commonly used in control theory textbooks in the control and design simulation toolkit, allowing for a natural transition from theoretical understanding to practical implementation. LabVIEW offers an open source interface for Arduino integration LINX toolbox, detailed in \cite{schwartz2015programming}, allowing affordable hardware platforms for their experiments. The process's cost-effectiveness facilitates wider accessibility and encourages hands-on learning experiences. 

The paper also presents an application of classical control techniques, mentioned in \cite{ogata2010modern}, including PID and Lead-Lag controllers, for a magnetic levitation system with an elastic beam. These choices are motivated by the Wide Application and Understanding of PID and Lead-Lag controllers as they represent fundamental control strategies widely used in various engineering domains. Their inclusion allows engineers to go through a well-established concepts with proven effectiveness, facilitating a solid foundation in control engineering principles, Moreover, these methods offer satisfactory performance for the chosen system, providing students with a practical demonstration of achievable control outcomes using well-understood techniques.

Although looking towards the future, This paper lays the groundwork for further exploration into advanced control approaches, including magnetic bearing and active magnetic bearing, described in \cite{shafai1994magnetic} and electromagnetic suspension applications such as MagLev trains, from \cite{sibilska2016use}, and other applications listed in \cite{yaghoubi2013most, han2016magnetic, app13020740}. MagLev offers promising solutions within the industrial sector and paves the way for investigating its integration within LabVIEW's framework for future educational endeavours.

\section{Hardware Setup}
Achieving operational efficiency in this system with heightened sensitivity necessitates a meticulous selection of components assuring precision in state measurement and accurate execution of desired actuator actions demands thoughtful consideration. The system involves the discerning choice of drivers for actuator control, in conjunction with microcontrollers to facilitate signal transmission and reception, choosing fundamental aspects of the system, mass, length of the cantilever, and the force exerted by the electromagnetic actuator. In this section, the selections made to develop the prototype are discussed.

\subsection{Cantilever Beam}
\begin{figure}[tb]
    \centering
    \includegraphics[width=0.6\textwidth]{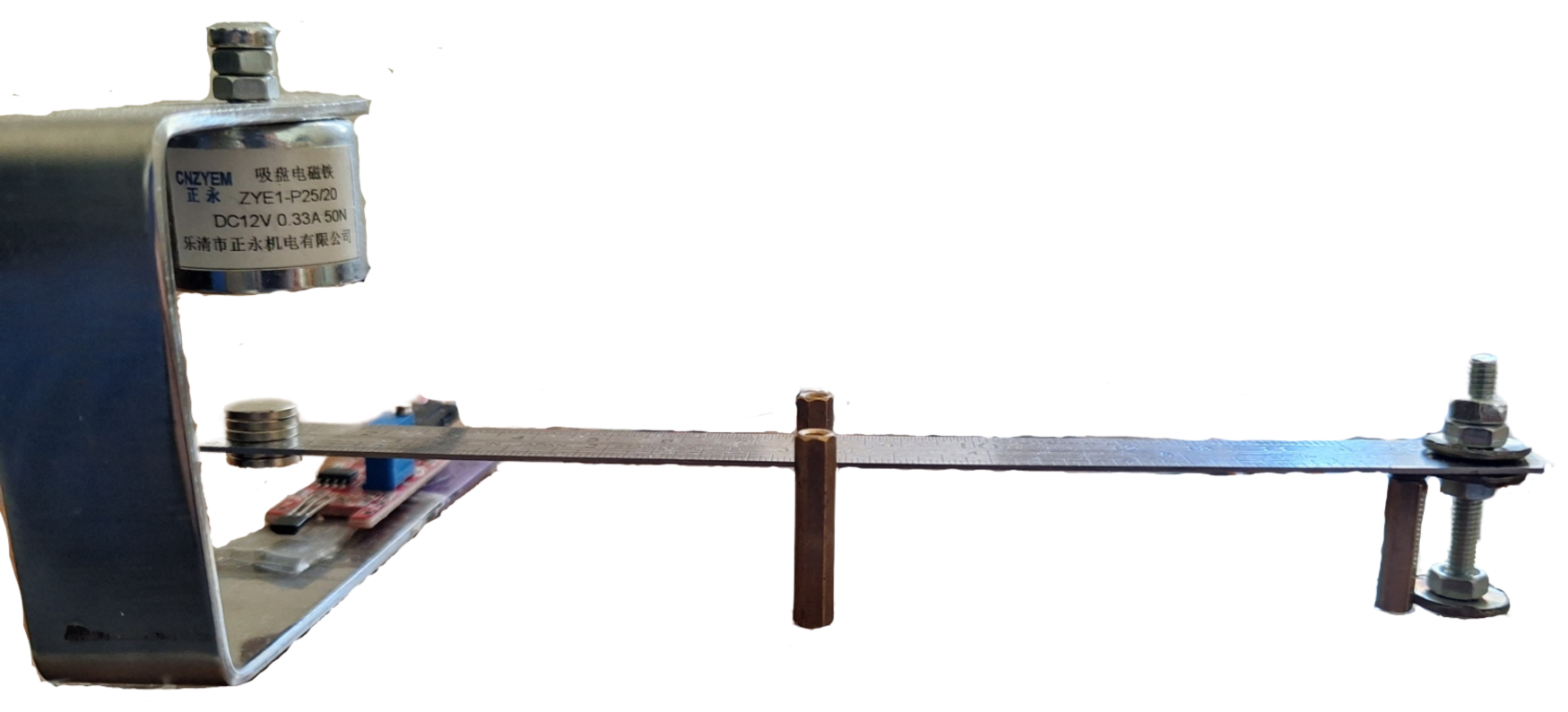}
    \caption{Cantilever beam fixation}
    \label{fig:cantilever_magnet}
\end{figure}

The initial aspect of consideration pertains to the selection of the cantilever. An "Endo Keiki" Stainless Steel Ruler shown in Fig. \ref{fig:cantilever_magnet} was chosen for this purpose. The cantilever is strategically equipped with $N35$ permanent magnets at the end tip of the beam, to interact with the electromagnet situated above the cantilever, and for sensing purposes, located beneath the cantilever, the sensor measures the position of these magnets, indicating the elevation of the cantilever.

\subsection{Electromagnet Actuator}
The chosen electromagnet is the "JF-XP$2520$ $12V$ DC" shown in Fig. \ref{fig:Top_system}-E holding force of up to $50N$. However, its availability and cost-effectiveness were more of a cause for choosing the electromagnet. The electromagnet is securely affixed to a stainless steel C-shaped stand, maintaining a consistent $35mm$ elevation from the sensor below as shown in Fig. \ref{fig:cantilever_magnet}. The space between the magnet and the sensor provides room for the cantilever, which can be adjusted in height. The electromagnet's magnetic field is bidirectional and controlled, and its force can be controlled for either pushing or pulling through a driver.

\subsection{Sensor}\label{sec:design/sensor}
\begin{figure}[h]
    \centering
    \includegraphics[width=0.7\textwidth]{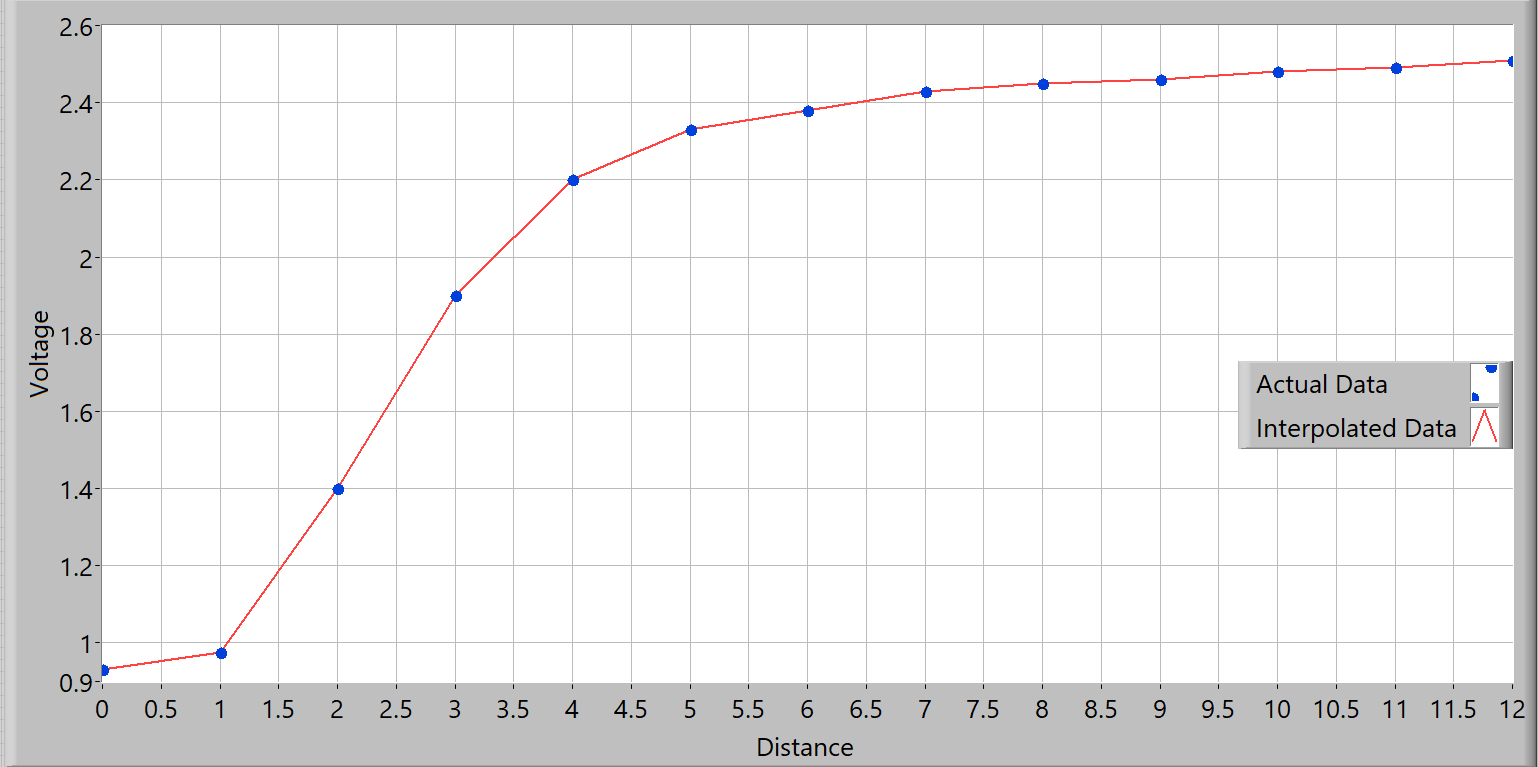}
    \caption{Sensor Calibration Interpolation}
    \label{fig:interpolation_graph}
\end{figure}

Distance sensor was quite a challenge, the Linear Hall Magnetic Sensor "KY-$024$" shown in Fig. \ref{fig:Top_system}-D with its working principle explained in \cite{sibilska2016use} and \cite{ramsden2011hall}, it detects the elevation of the magnet affixed to the cantilever with a very high sensitivity to small changes $\Delta V_{out} = 1.8 mV/G$, and after the AD converter the resolution is $0.02V$. For more reliability, a calibration process is done by taking numerous measurements. Subsequently, these measurements were interpolated, \cite{burden2011numerical}, and graphically represented as shown in Fig. \ref{fig:interpolation_graph}.

% \begin{table}[h!]
% \centering
% \begin{tabular}{||c | c c c c c c c ||} 
%  \hline
%  $y$ & 0 & 0.25 & 0.5 & 0.75 & 1.0 & 1.25 & 1.5 \\  
%  $V$ & 0.93 & 0.976 & 1.4 & 1.9 & 2.2 & 2.33 & 2.38 \\ [1ex]
%  \hline
%  \hline
%  $y$ & 1.75 & 2.0 & 2.25 & 2.5 & 2.75 & 3.0 & \\
%  $V$ & 2.43 & 2.45 & 2.46 & 2.48 & 2.49 & 2.51 & \\ [1ex]
%  \hline
% \end{tabular}
% \caption{Sensor calibration data}
% \label{table: calibration}
% \end{table}

%y = 0    0.2500    0.5000    0.7500    1.0000    1.2500    1.5000    1.7500    2.0000    2.2500    2.5000    2.7500    3.0000

%V = 0.9300    0.9760    1.4000    1.9000    2.2000    2.3300    2.3800    2.4300    2.4500    2.4600    2.4800    2.4900    2.5100

\subsection{Modules and Microcontroller}
\begin{figure}[h]
    \centering
    \includegraphics[width=0.6\textwidth]{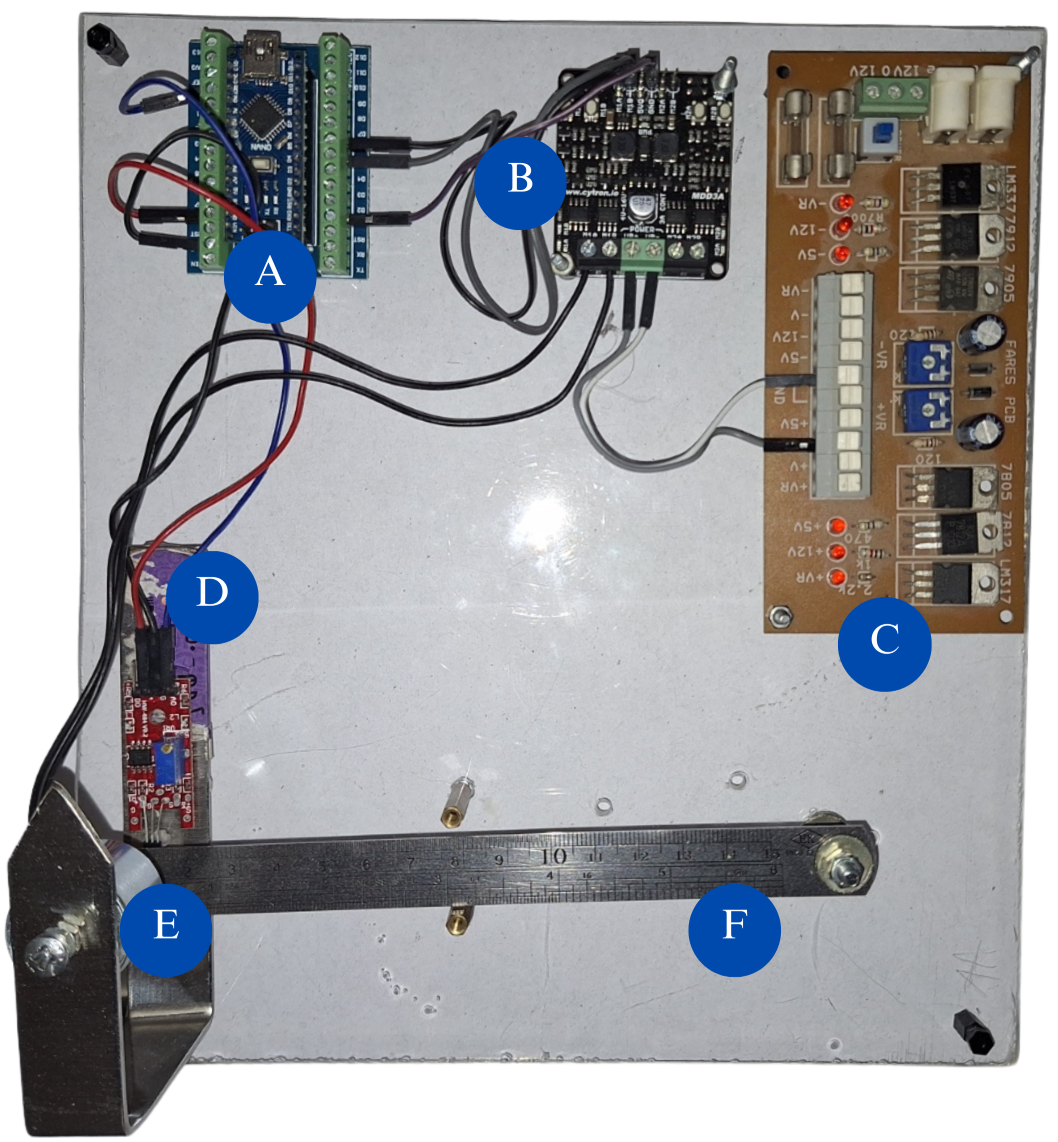}
    \caption{System Components: A-Microcontroller Arduino Nano, B-Motor Driver, C-Power Supply, D-Hall Effect Sensor, E-Electromagnet, F-Cantilever Beam}
    \label{fig:Top_system}
\end{figure}

The selection of the Arduino Nano shown in Fig. \ref{fig:Top_system}-A as our controller is underpinned by its rapid processing capabilities, ensuring minimal delays in managing control and measurement signals with baud rate set to $9600$. Furthermore, its seamless interfacing with LabVIEW makes it optimal for a small prototype.

For the motor driver, The 3Amp 4V-16V DC Motor Driver Cytron MDD3A shown in Fig. \ref{fig:Top_system}-B is opted. A low-cost, efficient component to supply power to the electromagnet, coupled with its demonstrated reliability and rapid responsiveness, aligning effectively with the demands of our system.

Lastly, as shown in Fig. \ref{fig:Top_system}-C a power supply module providing a stable $12V$ was incorporated to ensure the consistent and reliable operation of the entire system.

\subsection{LabVIEW VIs}
\begin{figure}[ht]
    \centering
    \includegraphics[width=0.8\textwidth]{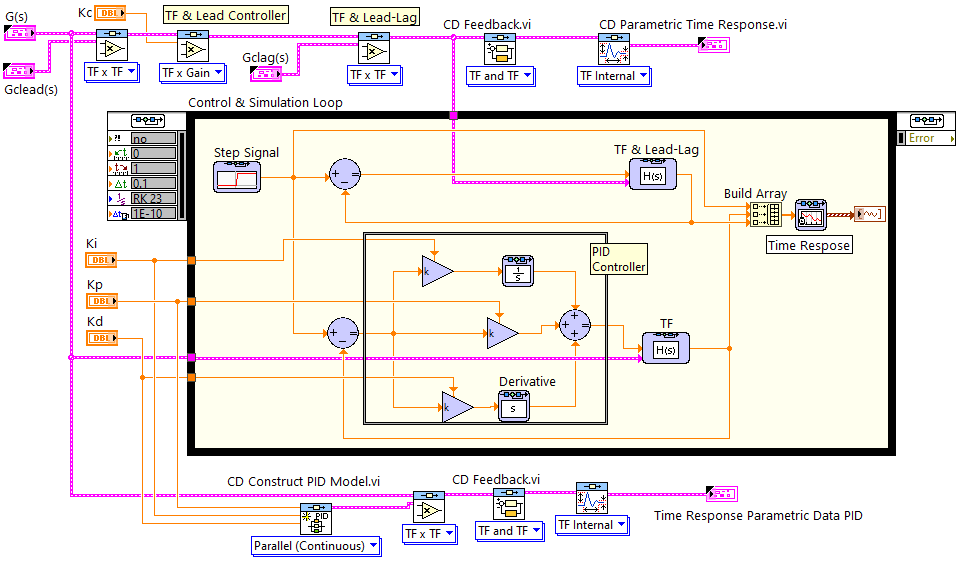}
    \caption{LabVIEW Block Diagram of System Simulation}
    \label{fig: vi sim}
\end{figure}

Building the virtual instrument VI of LabVIEW is considerably the key to the whole experiment, as illustrated in \cite{Bishop2015, modernControlSystemsLabVIEW}. Where the laws and theories can be applied to a prototype and compare the results to the model simulation to assess how accurate the modeling of the system was and choose which controller is more optimized. Fig. \ref{fig: vi sim} illustrates the control and design simulation toolkit usage to design the controller on an approximated transfer function of the system and obtain the time response from the simulation loop with the ability to change the design of Lead, Lag, and PID gain parameters.

\begin{figure}[ht]
    \centering
    \includegraphics[width=0.8\textwidth]{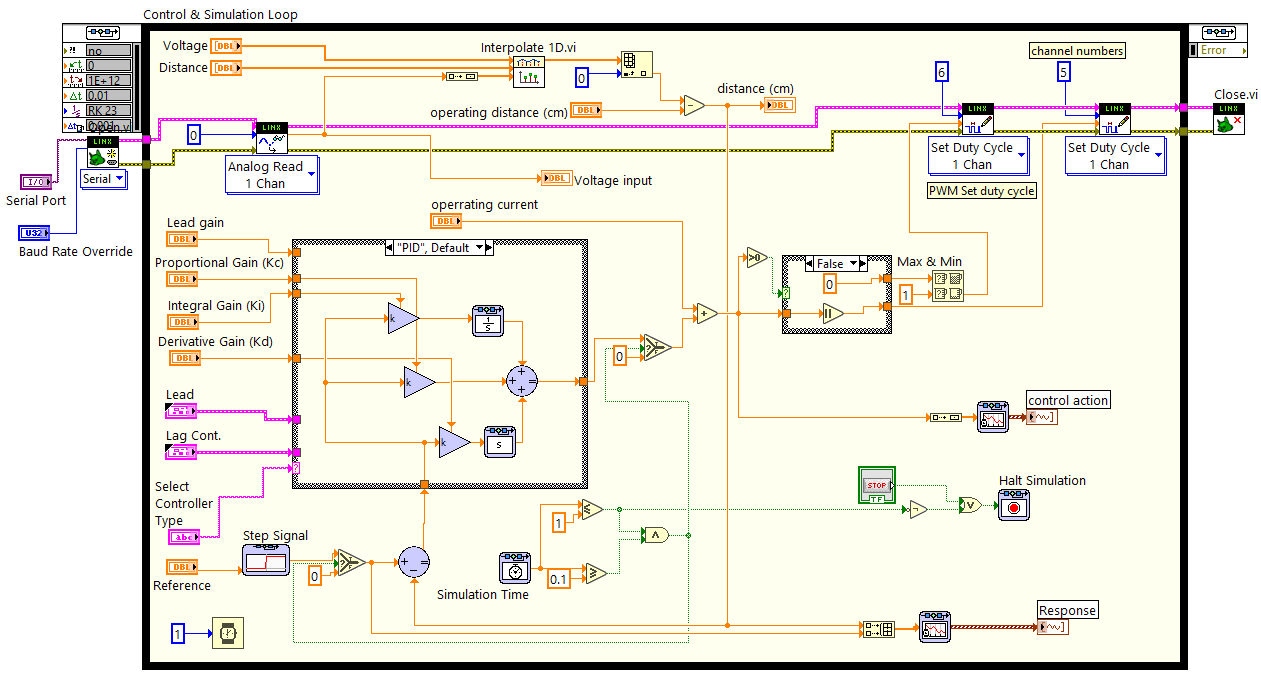}
    \caption{LabVIEW Block Diagram of Real Time Implementation using Arduino LINX-VIs}
    \label{fig: vi real}
\end{figure}

However, the LINX toolkit allows us to interface the microcontroller, clarified in \cite{schwartz2015programming}, either reading the voltage from the sensor or sending a pulse width modulation signal to the motor driver. Moreover, it enables the user to choose which controller to apply on the prototype, define its parameters, and visualize distance and control action, Fig. \ref{fig: vi real}. 

\section{System Modeling}
\begin{figure}[h]
    \centering
    \includegraphics[width=0.6\textwidth]{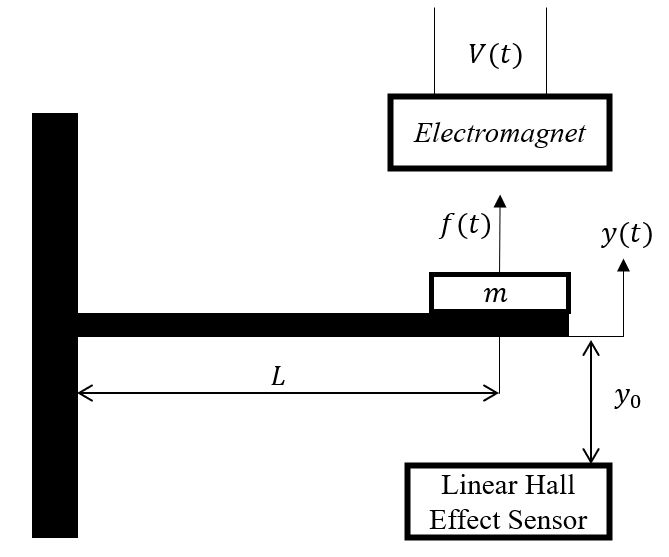}
    \caption{Magnetic Cantilever beam System}
    \label{fig:sytem}
\end{figure}

The system to be modeled -illustrated in Fig. \ref{fig:sytem}- can be described as a cantilever beam with a magnet attached at one end with mass $m$, and an electromagnet positioned nearby above, capable of attracting or repelling the magnet when a voltage $V(t)$ is applied to its terminals, and a sensor used to measure the distance $y(t)$ between the magnet and the electromagnet with an offset $y_0$.

In the process of system modeling for the described cantilever beam with a magnet and an electromagnet system, it is essential to establish a mathematical representation that captures the dynamic interactions and responses of the components involved. However, it is essential to make some approximations and assumptions to make the model easier to define and model.

\begin{figure}[h!]
\centering
\begin{subfigure}{0.35\textwidth}
    \includegraphics[width=\textwidth]{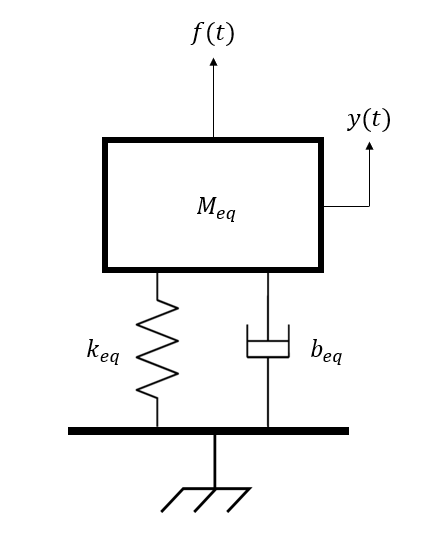}
    \caption{}
    \label{fig:modeled system-b}
\end{subfigure}
\begin{subfigure}{0.35\textwidth}
    \includegraphics[width=\textwidth]{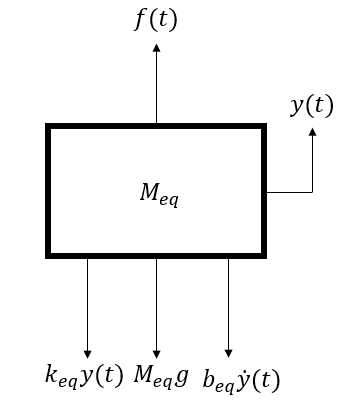}
    \caption{}
    \label{fig:modeled system-a}
\end{subfigure}   
\caption{modeled system as a (a) schematic (b) free body diagram}
\label{fig:modeled system}
\end{figure}

\subsection{Assumptions for Magnetic Levitation Model}
In our model, it is assumed that:
\begin{itemize}
    \item The magnet and the electromagnet are far apart, no forces are acting between them. Allowing us to focus on the interactions that occur when the electromagnet is activated.
    \item An operating point as a reference for our analysis, where the system operates around this specific condition. By focusing on small deflections, aiming to ensure that our analysis remains within the linear range of the system, simplifying the mathematical representation.
\end{itemize}

% \vspace{2mm}
\subsection{Cantilever Beam}
Primarily, the mechanical characteristics of the cantilever beam require representation through principles of structural mechanics. Parameters such as beam length $L$, elastic modulus $E$, and geometric shape (moment of inertia $I$) contribute to the determination of stiffness $k_eq$, equivalent mass $M_eq$, and damping coefficient $b_eq$. Given parameters for the cantilever beam system; Cantilever length $L = 150mm$,  Cantilever mass $m_c = 0.02Kg$, Magnet mass $m = 0.0075Kg$, Cantilever width $W = 16mm$, Cantilever thickness $t = 0.7mm$, Cantilever modulus of elasticity $E = 195GPa$.

%\begin{figure}[h!]
%\centering
%\begin{subfigure}{0.17\textwidth}
%    \includegraphics[width=\textwidth]{Figures/scheamtic.png}
%    \caption{}
%    \label{fig:modeled system-b}
%\end{subfigure}
%\begin{subfigure}{0.17\textwidth}
%    \includegraphics[width=\textwidth]{Figures/free body diagram.png}
%    \caption{}
%    \label{fig:modeled system-a}
%\end{subfigure}   
%\caption{modeled system as a (a) schematic (b) free body diagram}
%\label{fig:modeled system}
%\end{figure}

Using some calculations, the given system can be converted into a mass-spring-damper system with $M_{eq}$, $k_{eq}$, $b_{eq}$ that is easy to formulate its motion model from the free body diagram, the equation of motion of the mass-spring-damper system can be expresses as:
\begin{equation}\label{eq: motion}
    M_{eq} \ddot y + b_{eq} \dot y + k_{eq} y= f(t) - M_{eq}g
\end{equation}

From equation \ref{eq: motion}, The equivalent mass, damping and spring constants are to be calculated from the geometric of the cantilever beam and magnet and using the equations from \cite{cochin1997analysis}, Appendix 2, table 2.1.4-1, table 2.2.3-2:
\begin{equation}
    M_{eq} = m + 0.25m_c = 0.0123Kg
\end{equation}
\begin{equation}
    k_{eq} = \frac{3EI_{zz}}{L^3} = 74N/m
\end{equation}
As the damping coefficient is hard to calculate and depends on the medium and the beam, it is assumed with a small value to make the model work near the realistic response, by intuition: 
\begin{equation}
    b_{eq} = 0.1Ns/m
\end{equation}
Now, returning to the motion equation (Eq.\ref{eq: motion}) and transforming it into Laplace domain:
\begin{equation}\label{eq: motionL}
\begin{split}
    M_{eq}S^2Y(s) + b_{eq}SY(s) + k_{eq}Y(s) &= F(s) - M_{eq}g\\
\end{split}
\end{equation}
By making the substitution $F(0) =  M_{eq}g$, so the transfer function will be as follows:
\begin{equation}\label{eq: transferf}
\begin{split}
    \frac{Y(s)}{F(s)} = \frac{1/M_{eq}}{S^2 + \frac{b_{eq}}{M_{eq}}S + \frac{k_{eq}}{M_{eq}}}
\end{split}
\end{equation}

\subsection{Electromagnet}
Thirdly, the electromagnet is subjected to an electrical modeling approach. Within the framework of this electrical model, a linear assumption is incorporated to relate the applied voltage to the magnetic force. This assumption is justified by the consideration of a small time constant within the electrical circuit. By assuming linearity, a more straightforward analysis of the system's behavior is achieved in response to applied voltage variations, linear constant $K_v = 4.167$.
\begin{equation}\label{eq: FV}
    V(s) = K_v F(s) = 4.167 F(s) 
\end{equation}

\subsection{Sensor}
Calibration procedures and empirical data facilitate the establishment of a functional relationship between the sensor's output voltage and the displacement between the magnet and the electromagnet as illustrated in section \ref{sec:design/sensor}, so the sensor can be considered as a unity feedback for the system.

\begin{figure}[h]
    \centering
    \includegraphics[width=0.75\textwidth]{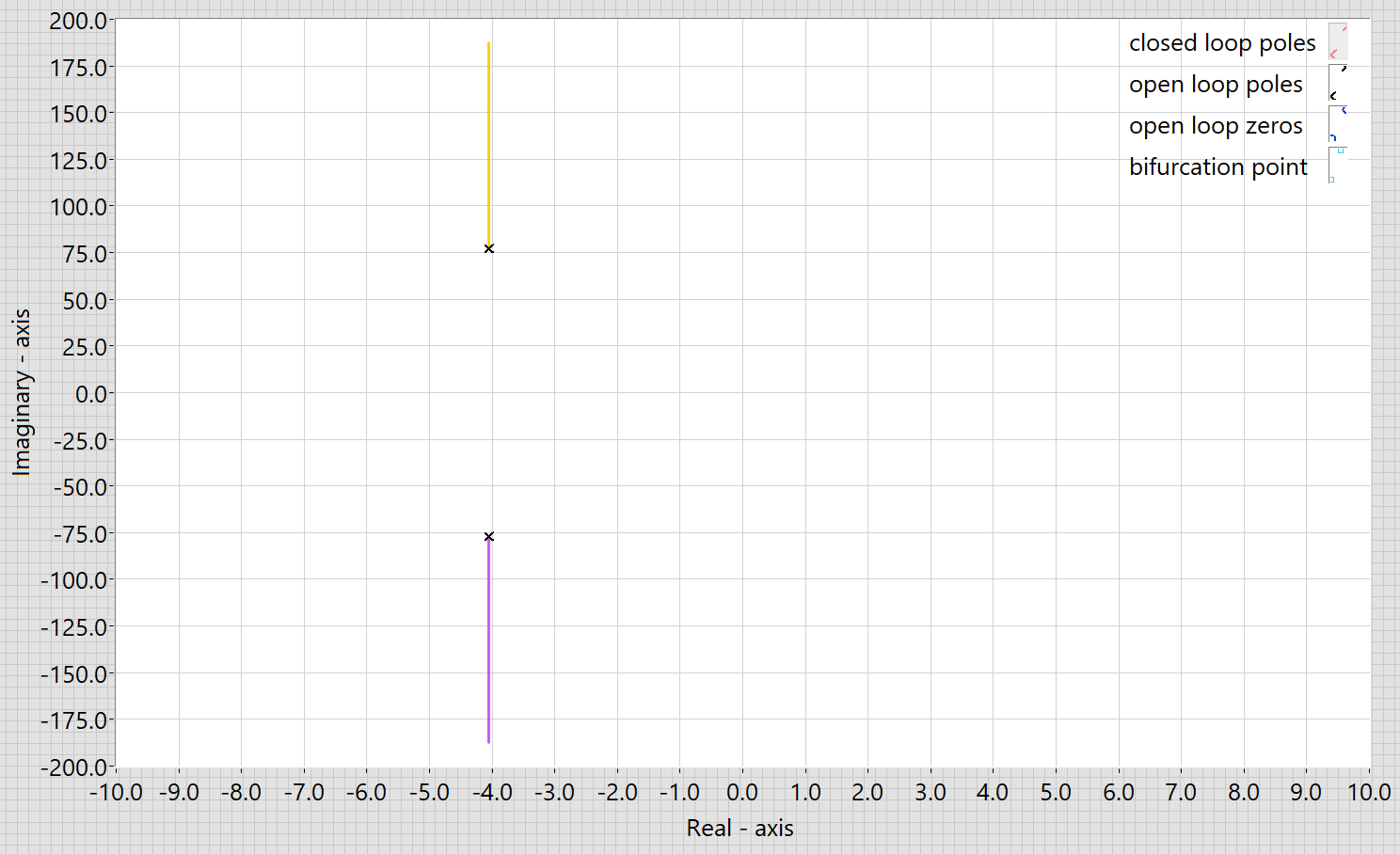}
    \caption{Root Locus - Uncompensated system - using LabVIEW}
    \label{fig:rootlocusofsytem}
\end{figure}

Finally, from equations \ref{eq: transferf} and \ref{eq: FV}, the final transfer function describing the whole system is achieved:
\begin{equation}\label{eq: transfer functoin}
    \frac{Y(s)}{V(s)} = \frac{K_v/M_{eq}}{S^2 + \frac{b_{eq}}{M_{eq}}S + \frac{k_{eq}}{M_{eq}}} = \frac{337.5}{S^2 + 8.1S + 5994}
\end{equation}
This gives us a root locus of the second order system as shown in Fig. \ref{fig:rootlocusofsytem}, and that makes the system a stable system as all poles are in the negative side of the real axis with poles at $s= -4.05 \pm 77.31j$

\section{Controller Design}
The controller design for the magnetic cantilever beam system discusses the optimization of two important classic control algorithms: A. Proportional-Integral-Derivative (PID) B. the Lead-Lag compensator. Optimizing these controllers will ensure the dynamic behavior of the system and stability of the system, defining the response of the cantilever beam to track reference or reject disturbances.

\begin{figure}[h]
    \centering
    \includegraphics[width=0.65\textwidth]{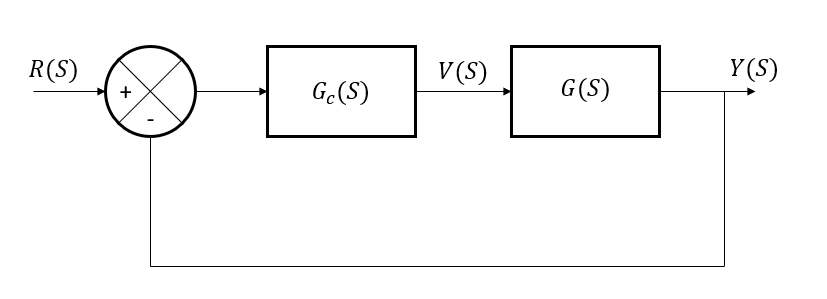}
    \caption{Block diagram of system with compensator}
    \label{fig: blockdiagram}
\end{figure}

\subsection{PID Control}
The PID controller, a fundamental algorithm in control theory, has been extensively explored in the literature, including studies by \cite{shaban2013proportional, Nada2014}, \cite{alsayed2018fuzzy}. It aims to optimize the magnetic cantilever beam's response to reject disturbances, particularly those induced by changes in the magnetic field. The formula used is the parallel continuous PID gains given by:
\begin{equation}\label{eq: Gc PID}
\begin{split}
    G_c &= K_p + \frac{K_i}{s} + K_d s\\
\end{split}
\end{equation}

Designing a PID controller is necessary to achieve a smooth response, disturbance rejection, and reference tracking, using the PID tuner algorithm from \cite{aastrom2006advanced} can ensure a tuned response for the system plant with the block diagram shown in Fig. \ref{fig: blockdiagram}.
The tuned response is set to have a small settling time, low overshoot percentage and limited output voltage between $-12V : 12V$ and the tuner sets the gains to; $K_p = 6.55$, $K_i = 149.36$, $K_d = 0.022$.
by substituting in PID transfer function equation \ref{eq: Gc PID} that gives us the controller transfer function $G_c$ of:
\begin{equation}\label{eq: Gc PID sub}
    G_c = 6.55 + \frac{149.36}{s} + 0.022 s
\end{equation}

\begin{figure}[h]
    \centering
    \includegraphics[width=0.8\textwidth]{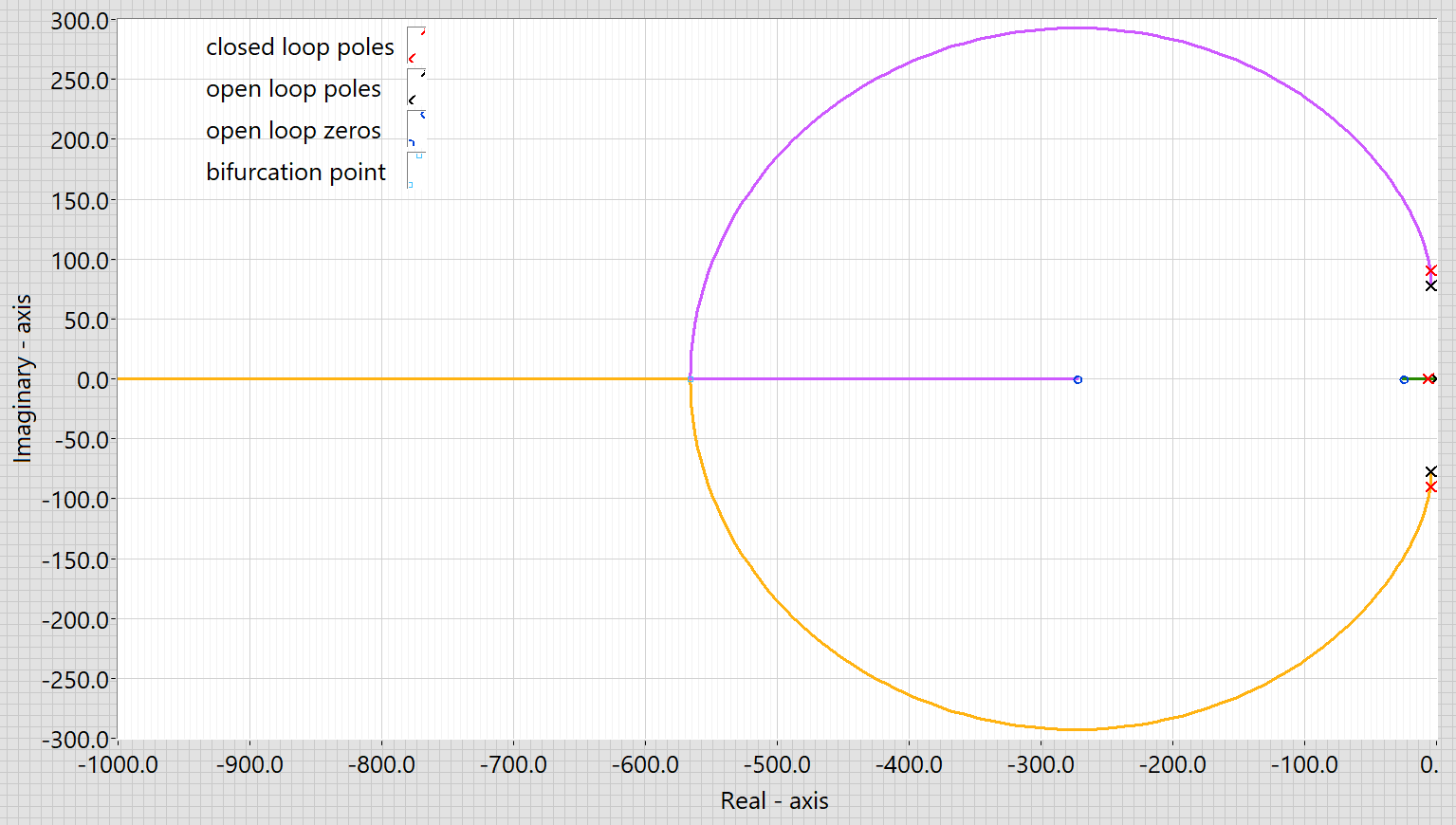}
    \caption{Root Locus of compensated system - PID }
    \label{fig:rootlocuspid}
\end{figure}

Investigating the root locus (Fig. \ref{fig:rootlocuspid}) of the compensated system, it's fair to say that the closed loop poles of the root locus ensure that the system is stable. 

\subsection{Lead-Lag Control}
Concurrently, the Lead-Lag controller introduces a phase lead at specific frequencies and phase lag at others, explained in \cite{ogata2010modern}, enhancing the overall responsiveness of the system. Lead-Lag controller transfer function contains a pole and zero for the lead the zero leads the pole to get a specific dynamic behavior and the lag has a pole and zero with a zero close to the origin to minimize the error, the transfer function is as follows:

\begin{equation}\label{eq: Gc leadlag}
    G_c = K_c \frac{s + \frac{1}{\tau_{lead}}}{s + \frac{1}{\alpha\tau_{lead}}} \frac{s + \frac{1}{\tau_{lag}}}{s + \frac{1}{\beta\tau_{lag}}}
\end{equation}
where $1 < \alpha < 0$ and $\beta > 1$. The Lead-Lag design process is made by defining desired poles of $s= -200 \pm 450j$ targeting for an objective of an overshoot $M_p = 0.25$ and settling time $T_s = 10ms$ designed by the traditional method, \cite{ogata2010modern}, to get the presented transfer function for the compensator:

\begin{equation}\label{eq: Gc leadlag sub}
    G_c = 11 \frac{s + 180}{s + 750} \frac{s + 0.001}{s + 0.0001}
\end{equation}
Where ; $K_c = 11$, $\tau_{lead} = 0.00556$, $\alpha = 0.24$, $\tau_{lag} = 1000$, $\beta = 10$.

\begin{figure}[h]
    \centering
    \includegraphics[width=0.8\textwidth]{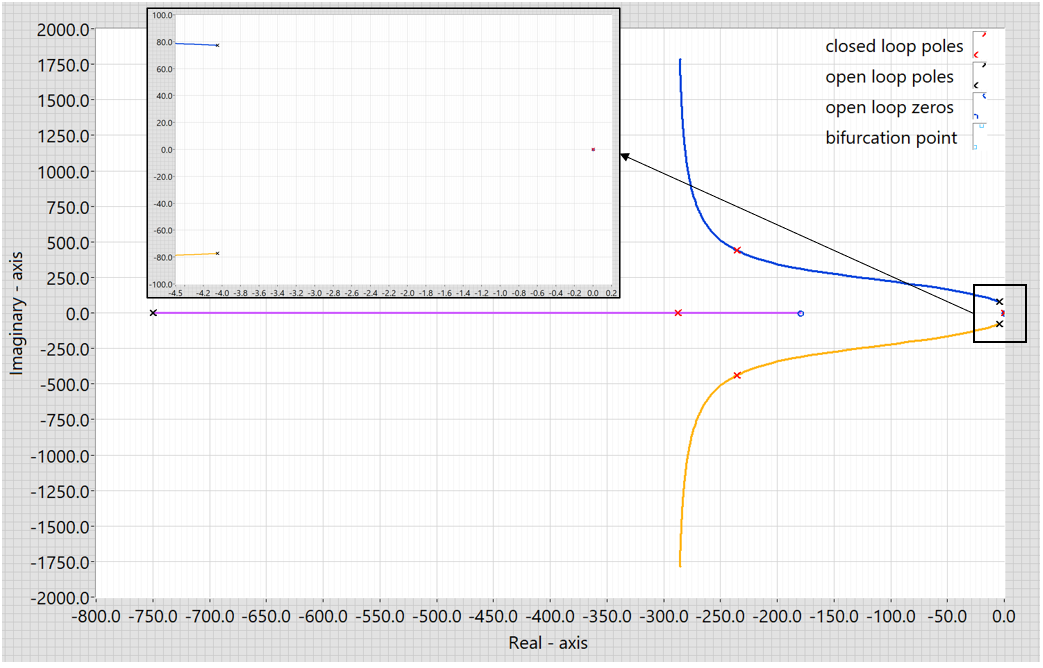}
    \caption{Root Locus of Compensated system - Lead Lag}
    \label{fig: leadlag rootlocus}
\end{figure}

The compensated system with the transfer function of equation \ref{eq: Gc leadlag sub} can be modeled into the root locus (Fig. \ref{fig: leadlag rootlocus}), anticipating the closed loop poles at the gain $K_c = 11$, which it should gives the targeted overshoot and settling time.

\begin{figure}[h]
    \centering
    \includegraphics[width=0.8\textwidth]{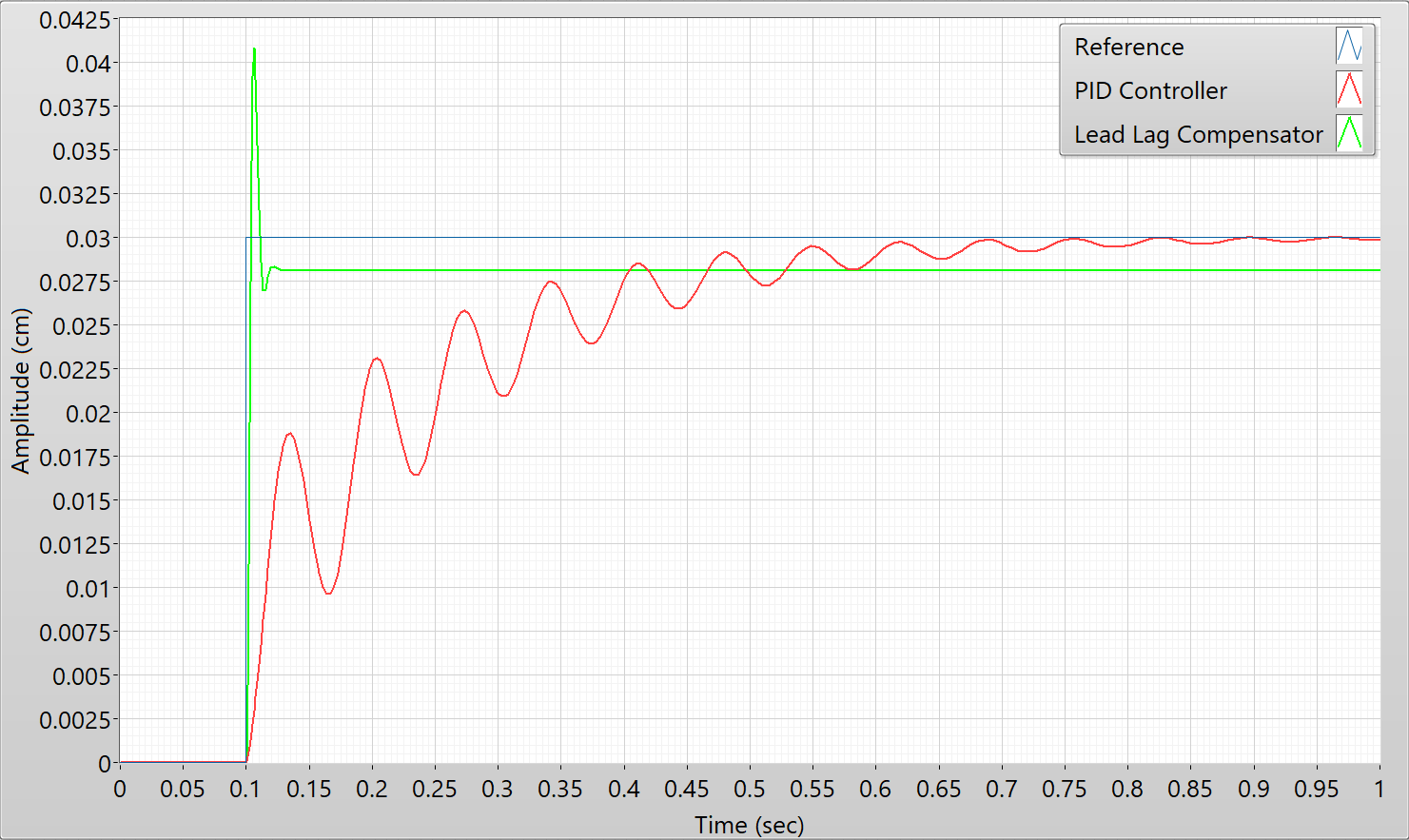}
    \caption{Simulation response of PID vs Lead-Lag controllers}
    \label{fig: sim_time_res}
\end{figure}
% \vspace{3mm}
% discussion of the simulation results pid vs lead lag
In Fig. \ref{fig: sim_time_res}, the red solid line shows the PID controller response to the step input (blue solid line), and the green solid line is the Lead-Lag compensator response to the same step input. Also, from the LabVIEW time parameter calculator, the PID response is $T_{s} = 0.776578s$, $M_{p} = 0.116312\%$, $e_{ss} = 0$, the Lead-Lag response is $T_{s}$ = $0.025s$, $M_{p}$ = $37.576$\%, $e_{ss}$ = $0.0025$.\\
PID behavior can be explained by the fact that the closed loop poles from the root locus shown in Fig. \ref{fig:rootlocuspid} define the system as a third-order response with a dominant first-order response and some oscillations to reach a steady state error.
On the other hand, it's noted that the Lead-Lag response is as predicted before with a little high overshoot due to the sensitivity of the system, however, showing a small settling time compared to the PID response.

\section{Results}
\subsection{Solver Optimization}\label{sec: solver}
To apply dynamic system modeling, solving differential equations is necessary to get the time response of the simulation model and the real-time to apply the compensator function. LabVIEW Control Design and Simulation toolkit offers ordinary differential equation (ODE) solvers for simulation. In specific systems, such as ours, the Euler solver's computation time induces continuous oscillations, and fluctuating actions.\\
To tackle this problem, modifications were introduced to enhance the solver's response by applying Runge-Kutta $23$ solver with variable step time (range $10ms : 100ms$, tolerance $0.1$, sampling time $0.01s$), which notably improved the system's response, although the solver with smaller step time will take too long to compute the output losing the real-time ability of the system. 

\begin{figure}[h]
    \centering
    \includegraphics[width=0.7\textwidth]{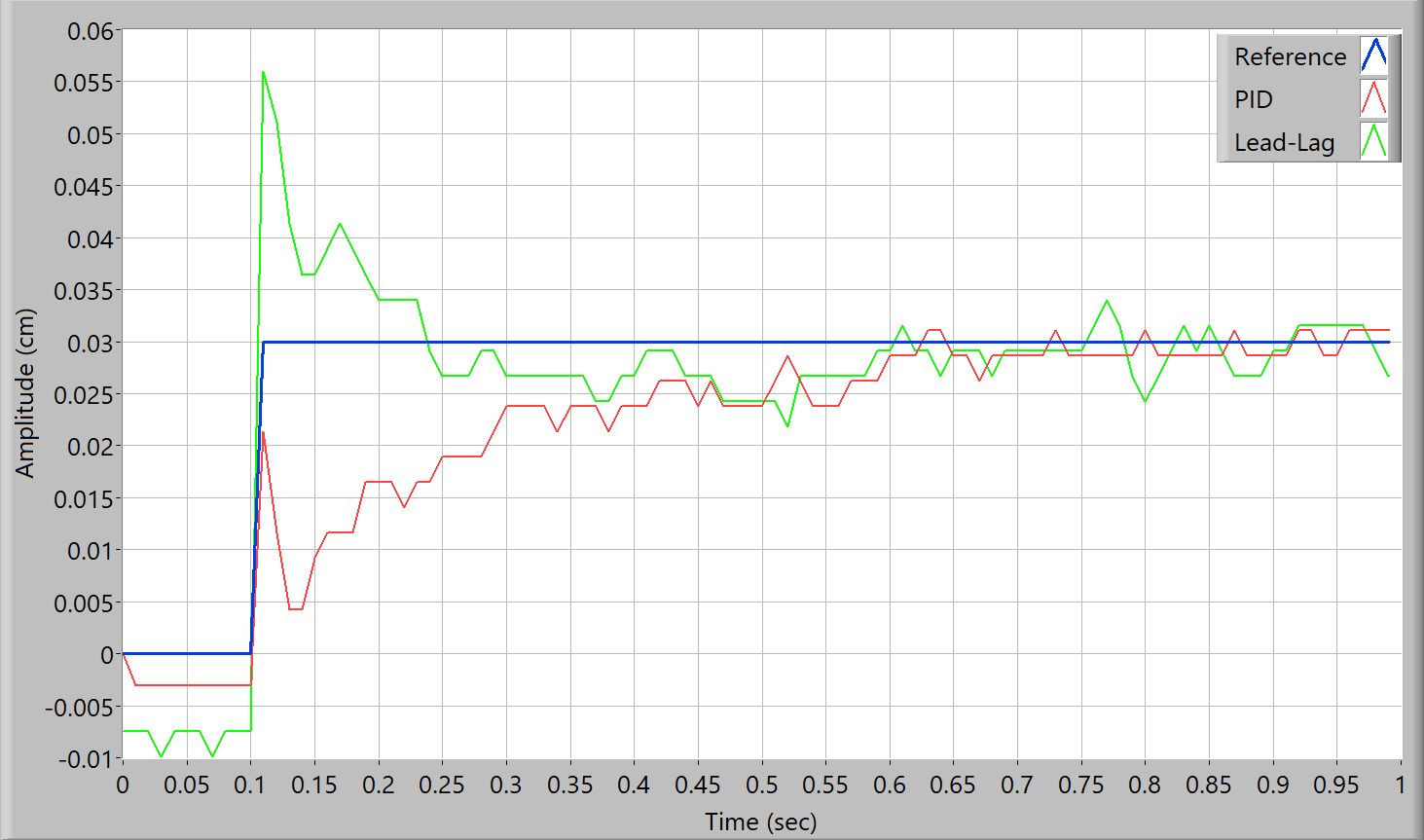}
    \caption{Real Time Response of PID vs Lead-Lag Controllers}
    \label{fig:real_time_res}
\end{figure}

\begin{figure}[h]
    \centering
    \includegraphics[width=0.7\textwidth]{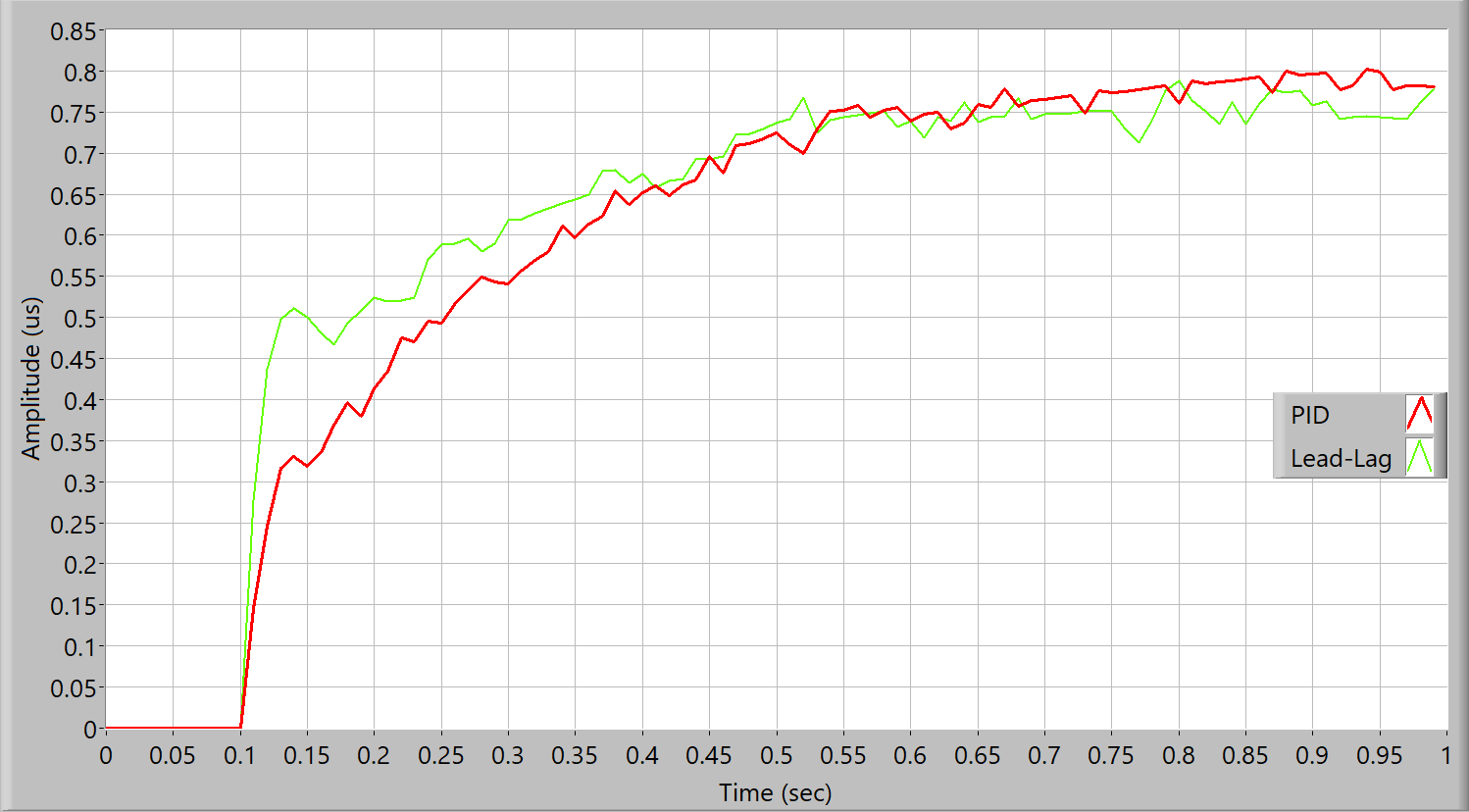}
    \caption{Action amplitude of PID  vs Lead-Lag Controller}
    \label{fig:action_time_res}
\end{figure}

%%% -------------------------------------- Edites till here

\subsection{Comparison}

\begin{table}[h!]
\centering
\caption{PID data}
\begin{tabular}{||c | c | c||} 
 \hline
  & Simulation & Real \\ [1ex]
 \hline
 $T_{s}$ & $0.776578s$ & $0.6s$ \\  [1ex]
 \hline
 $M_{p}$ & $0.116312$\% & $0$\% \\  [1ex]
 \hline
 $e_{ss}$ & $0$ & $0$ \\  [1ex]
 \hline
\end{tabular}
\label{table:PID_Parameters}
\end{table}

\begin{table}[h!]
\centering
\caption{Lead-Lag data}
\begin{tabular}{||c | c | c||} 
 \hline
  & Simulation & Real \\ [1ex]
 \hline
 $T_{s}$ & $0.025s$ & $0.15s$\\  [1ex]
 \hline 
 $M_{p}$ & $37.576$\% & $90$\%  \\  [1ex]
 \hline
 $e_{ss}$ & $0.0025$ & $0$ \\  [1ex]
 \hline
\end{tabular}
\label{table:Lead-Lag_Parameters}
\end{table}

Lead-Lag controller time response from Fig. \ref{fig:real_time_res} (green solid line) and table. \ref{table:Lead-Lag_Parameters} from LabVIEW time parameter calculator, indicates a similar behavior to that of the simulation in Fig. \ref{fig: sim_time_res}, with a settling time slightly longer than the simulation settling time, as indicated in table.\ref{table:Lead-Lag_Parameters}. Even so, the overshoot exceeded the simulation overshoot by far arguably because of the sensitive structure that is stimulated by small forces. Steady-state error remained nearly zero verifying the Lag compensator effect to the system.
\\
In contrast, PID controller proves advantageous by eliminating overshoot entirely, along with a settling time of 0.6 seconds showing high similarity to the simulation results, Fig. \ref{fig: sim_time_res} and Table. \ref{table:PID_Parameters}. Compared to Lead-Lag compensator, Lead-Lag has a shorter settling time of 0.15 seconds, but with aggressive overshoot, and it's fair to say that both have approximately 0 steady state error.
\\
Also, control action is an aspect that must be taken into consideration due to the limitation of voltage supply to the electromagnet. In the comparison of control action amplitude Fig. \ref{fig:action_time_res}, PID shows a smooth gradual response with small fluctuations due to the discrete time step of the solver, section \ref{sec: solver}. Where Lead-Lag is seen to have a high amplitude in the beginning causing the high overshoot response, but stabilizes and reaches the same amplitude as PID.

\section{Conclusion}
The paper work reveals that the LabVIEW G-programming language provides a user-friendly interface that positively contributes to the understanding of control theory, and witnessing the effect of optimizing and tuning parameters on the time response of the system. It is experiencing the influence of selecting ODE solvers, the step size, and the required accuracy with respect to the real-time constraints. Moreover, it could verify the system modeling, approximations, simulation, and real-time implementation. It also facilitates the interface with low-cost microcontrollers through serial data transfer. In conclusion, LabVIEW enables the engineer to maximize system performance with an economical and low-priced control solution. The paper presents the dynamic model of magnetic levitation system, the frequency and time analysis and control system design by utilizing the PID and Lead-Lag controllers. The real-time implementation through hardware-in-loop control system shows satisfactory results with respect to the desired performance.

\bibliographystyle{plain}
\bibliography{references}

\end{document}